\documentclass[%
 aip,
 jmp,
 amsmath,amssymb,
preprint,%
]{revtex4-1}

\usepackage{graphicx}
\usepackage{dcolumn}
\usepackage{multirow}
\usepackage{bm}

\input xy
\xyoption{all}

\newtheorem{definition}{Definition}[section]
\newtheorem{lemma}{Lemma}[section]
\newtheorem{proposition}{Proposition}[section]
\newtheorem{theorem}{Theorem}[section]

\newtheorem{remark}{Remark}[section]

\newcommand{\cqfd}{\hfill $\square$}
\begin{document}


\title[]{GROUP-ALGEBRAIC CHARACTERIZATION OF SPIN PARTICLES: SEMI-SIMPLICITY, SO(2N) STRUCTURE AND IWASAWA DECOMPOSITION}

\author{Mahouton Norbert Hounkonnou}
\email{(corresponding author) norbert.hounkonnou@cipma.uac.bj (with copy to hounkonnou@yahoo.fr)}
\author{Francis Atta Howard}%
 \email{hfrancisatta@ymail.com; Francis$_$atta$_$Howard@cipma.net}

\affiliation{University of Abomey-Calavi,\\
International Chair in Mathematical Physics and Applications (ICMPA--UNESCO Chair), 072 B.P. 50  Cotonou,   Benin Republic
}%


\author{Kangni Kinvi}
\email{kangnikinvi@yahoo.fr}
\affiliation{Universit\'{e} de Cocody, \\
	UFR- Mathematiques et Informatique, 
	22 BP. 1214, Abidjan 22, C\^{o}te d'Ivoire
}%


\date{\today}

\begin{abstract}
In this paper, we  focus on  the characterization of  Lie algebras of fermionic, bosonic and parastatistic operators  of  spin particles. We provide a method to construct a Lie group structure for the quantum spin particles. We  show the semi-simplicity of a quantum spin particle Lie algebra, and extend the results to the Lie group level. Besides, we perform the Iwasawa decomposition of  spin particles at both the Lie algebra and Lie group levels. Finally, we  investigate the coupling of  angular momenta of spin half particles, and give   a  general construction for such a study.
\end{abstract}

\maketitle


\section{Introduction}
\subsection{motivations}
Doran's work\cite{ref4}  showed that every linear transformation can be represented as a monomial of vectors in geometric algebra, every Lie algebra  as a bivector algebra, and every Lie group  as a spin group.  Schwinger's \cite{ref16} realization of $su(1,1)$ Lie algebra  with creation and annihilation operators was defined with spatial reference  in  the Pauli matrix representation \cite{ref22}. Several relations as well as connections were observed in spin particles such as fermionic, bosonic,  parastatistic Lie algebras, and in geometric algebras such as the Clifford algebra, Grassmannian algebra and so on\cite{ref17}.  Sobczyk \cite{ref17} also proved that the spin half particles can be represented by  geometric algebras. Palev \cite{ref15} highlighted that a semi-simple Lie algebra can be generated by the creation and annihilation operators. In all the above mentioned works, the classical groups such as  $B_{n}$ and $D_{n}$ play a crucial role in the spin particle Lie algebra. Moreover, several evidences from particle and theoretical physics   showed  the connection between quantum spin particle Lie algebra and Clifford algebra\cite{ref4}.
The spin of elementary particles  obeying the Fermi-Dirac statistics, the Bose-Einstein statistics,  the quantization  of parastatistics such as parafermions and parabosons also  gained much attention  in the literature\cite{ref3}, \cite{ref4}, \cite{ref16},\cite{ref17}, \cite{ref21}, \cite{ref22}.\\
In the opposite,    exhaustive investigations on spin particle creation and annihilation and their angular momentum in connection with Lie groups, Lie algebras, Clifford algebras, and their representations are still lacking. This study aims at fulfilling this gap. The Iwasawa decomposition,  introduced by the Japanese mathematician Kenkichi Iwasawa, generalizes the Gram-Schmidt orthogonality process in linear algebra\cite{ref9}.

Motivated by all the above mentioned works, we prove, in this paper,  that the spin particles admit a Lie group structure, show its connectedness and  semi-simplicity, and  construct the Iwasawa decomposition at both the Lie algebra and Lie group levels of the spin particles.

But before dealing with the main results, and as a matter of clarity in the development, let us briefly recall the main definitions, the known results,  and the appropriate notations useful in the sequel.

\subsection{Para-fermionic algebra}
Let $a_{1}^{\pm},...,a_{n}^{\pm}$ be the creation and annihilation operators for a system consisting of n-fermions with commutator relations\cite{ref6}:
\begin{equation}
[a_{i}^{-},a_{j}^{+}]=\delta_{ij}
\label{eq1}
\end{equation}

\begin{equation}
[a_{i}^{-},a_{j}^{-}]=[a_{i}^{+},a_{j}^{+}]=0,
\label{eq2}
\end{equation}
or, of n-parafermions with 
\begin{equation}
[[a_{i}^{-},a_{i}^{+}],a_{j}^{\pm}]= \pm 2\delta_{ij}a_{j}^{+},
\label{eq3}
\end{equation}
where
\begin{equation}
[X,Y]:=XY-YX.
\label{eq4}
\end{equation}
Let $T$ be the associative free algebra of $a_{i}$, $a_{j}$; $i,j\in N=\{1,2,...n\}$, and $I$ be the two sided ideal in $T$ generated by the relation ~\eqref{eq3}. The Quotient (factor algebra) 
\begin{equation}
Q=\dfrac{T}{I}
\label{eq5}
\end{equation}
is called para-Fermi algebra, for all $X,Y\in Q$. This is an infinite dimensional Lie algebra with respect to the bracket defined by the equation ~\eqref{eq4}.
\subsection{Semi-simple  Lie algebra generated by creation and annihilation operators}
In this subsection, we quickly  review main  Lie algebraic properties  retrieved from the work by Palev\cite{ref15}, which are  useful for our construction performed by sticking to the same notations.
\\
Let $g$ be a semi-simple Lie algebra generated by $n$ pairs $a_{1}^{\pm},...,a_{n}^{\pm}$ of creation and annihilation operators. The elements
\begin{gather*}
h_{i}=\dfrac{1}{2}[a_{i}^{-},a_{i}^{+}], i=1,...n
\end{gather*}
are contained in a Cartan subalgebra $H$ of $g$. The rank of $g\geq n$.
If the semi-simple Lie algebra $g$ of rank $n$ is generated by $n$ pairs of creation and annihilation operators, then, with respect to the basis of the Cartan subalgebra, the creation (resp. annihilation) operators are negative (resp. positive) root vectors.
The correspondence with their roots is:
\begin{gather*}
a_{i}^{\pm}\longleftrightarrow \pm h^{*i}.
\end{gather*}
where $\pm h^{*i}$ is a basis in the space dual to the Cartan subalgebra\cite{ref15}.

The semi-simple Lie algebra $g$ of rank $n$ is generated by $n$ pairs of creation and annihilation operators if and only if it contains a complete system $\Phi$ of roots orthogonal with respect to the Cartan-Killing form.
The semi-simple Lie algebra $g$ of rank $n$ is generated by $n$ pairs of creation and annihilation operators if and only if it is a direct sum of classical Lie algebras
\begin{gather*}
g=B_{m_{1}}\oplus...\oplus B_{m_{k}}
\end{gather*} 
where $m_{1}+...+m_{k}=n$. \\
The simple Lie algebra $g$ of rank $n$ is generated by $n$ pairs of creation and annihilation operators if and only if it is isomorphic to the classical Lie algebra $B_{n}$.\cite{ref15}
\\

To construct an example of a semi-simple Lie algebra,  we adapt Schwinger notation for the $su(1,1)$ Lie algebra: Let $a_{r}^{+}=(a_{+}^{+},a_{-}^{+})$ and $a_{r}=(a_{+},a_{-})$
be the spin creation and annihilation operators, which obey the following commutation relations:
\begin{enumerate}
	\item [] $[a_{r},a_{r^{\prime}}]=0;\;\;
	[a_{r},a_{r^{\prime}}^{+}]={\delta_{r}}_{r^{\prime}};\;\;\;
	[a_{r}^{+},a_{r^{\prime}}^{+}]=0 $
\end{enumerate}
and the number of spins and the resultant angular momentum($j=\frac{1}{2}$) are respectively given by; $n=\sum_{r}a_{r}^{+}a_{r}$ , $j=\sum_{{r}{r^{\prime}}}{a^{+}_{r}}\left(r\mid \dfrac{1}{2} \sigma \mid r^{\prime}\right)a_{r},$
where the creation and annihilation operators $a^{+}$ and $a$  may be defined as:
\begin{gather*}
a^{+}=\dfrac{1}{\sqrt{2}}\left(x-\dfrac{\partial}{\partial x}\right);\;\;\;
a=\dfrac{1}{\sqrt{2}}\left(x+\dfrac{\partial}{\partial x}\right),
\end{gather*}
which satisfy the  Schwringer\cite{ref16} $su(1,1)$ Lie algebra for the one-dimensional harmonic oscillator characterized by:
 $K_{+}=\dfrac{1}{2}a^{+}a^{+}$ , $K_{-}=\dfrac{1}{2}aa$, $K_{z}=\dfrac{1}{2}(a^{+}a + 1)$, $K_{\pm}=K_{x}\pm iK_{y},$ and  the Casimir operator  $C=-K_{x}^{2}-K_{y}^{2}+K_{z}^{2},$ 
where the operators $K_{+}$, $K_{-}$ and $K_{z}$ obey  the commutation relation:
\begin{gather*}
[K_{z},K_{\pm}]=\pm K_{z};\;\;\;
[K_{+},K_{-}]=-2 K_{z}.
\end{gather*}	
\begin{remark}
	The above $su(1,1)$ quasi-boson Lie algebra  is a semi-simple Lie algebra.
\end{remark}

\subsection{Lie algebra of spin group}
 Let now $m$ be an $n$-dimensional oriented real vector space with an inner product $<,>$. We define the Clifford algebra \cite{ref7} $Cl(m)$ over $m$ by the quotient ${T(m)}/{I}$, where $T(m)$ is a tensor algebra over $m$ and $I$ is the ideal generated by all elements $v\otimes v+<v,v> 1$, $v\in m$. The multiplication of $Cl(m)$ will be denoted by $x\cdot y$. Let $p:T(m)\longrightarrow Cl(m)$ be the canonical projection. Then, $Cl(m)$ is decomposed into the direct sum $Cl^{+}(m)\oplus Cl^{-}(m)$ of the p-images of the elements of even and odd degrees of $T(m)$, and $m$ is identified with the subspace of $Cl(m)$ through the projection $p$. Let ${e_{1},e_{2},\cdots,e_{n}}$ be an oriented orthonormal basis of $m$. The map: $e_{i2}\cdot e_{i2}\cdot \cdots \cdot e_{ip}\mapsto (-1)^{p}e_{ip}\cdot \cdots e_{i2}\cdot e_{i1}$ defines a linear map of $Cl(m)$ and the image of $x\in Cl(m)$ by this linear map is denoted by $\bar{x}$. The spin group is defined by:
\begin{equation}
Spin\thickspace(m)=\{{x\in Cl^{+}(m): x \thickspace is \thickspace invertible, \thickspace x\cdot m \cdot x^{-1}\subset m \thickspace and \thickspace x\cdot \bar{x}=1}\}.
\label{eq6}
\end{equation}	
Moreover, the subspace $spin\thickspace(m)$ of $Cl(m)$ spanned by $\{e_{i}\cdot e_{j}\}_{i<j}$ is a Lie algebra of $Spin\thickspace(m)$ in such a way that $\exp: spin\thickspace(m)\rightarrow Spin\thickspace(m)$ is just the restriction of the exponential map of the algebra $Cl(m)$ into $Cl(m)$. The differential $\dot{\pi}$ of $\pi$ is given by:
\begin{gather*}
\dot{\pi}(x)v=x\cdot v-v\cdot x,
\end{gather*}

for $x\in spin\thickspace(m)$ and $v\in m$.

\subsection{ Root system for semi-simple Lie group}
 Eugene Dynkin, based on a  geometric method of classifying all simple Lie groups,  proved that the semi-simple Lie group is determined by its system of simple roots. The root system of the group
$B_{n}$ is the group of orthogonal transfromations of a $(2n+1)$-dimensional complex vector space\cite{ref5} $L^{2n+1}$:
\begin{equation*} 
B_{n}, n\geq2; g=SO(2n+1, \mathbb{C}), 
K=SO(n, \mathbb{C})\times SO(n+1,\mathbb{C})
\label{eq7}
\end{equation*}
with $n=2m (n=2m+1)$,
\begin{equation}
\sum(B_{n})=\left\{\pm e_{p},\pm e_{p}\pm e_{q}\right\}^{n}_{p,q}=1 (p\neq q;e_{1},\cdots,e_{2n+1} \thickspace is\thickspace an \thickspace orthonormal \thickspace basis\thickspace of \thickspace R^{2n+1}),
\label{eq8}
\end{equation}
and $D_{n}$ is the group of orthogonal transfromations of a $(2n)$-dimensional complex vector space\cite{ref5} $L^{2n}:$
\begin{equation*} 
D_{n}, n\geq4; g=SO(2n, \mathbb{C}), 
K=SO(n, \mathbb{C})\times SO(n,\mathbb{C})
\label{eq9}
\end{equation*}
with $n=2m (n=2m+1)$,
\begin{equation}
\sum(D_{n})=\left\{\pm e_{p}\pm e_{q}\right\}^{n}_{p,q}=1 (p\neq q; e_{1},\cdots,e_{2n} \thickspace is\thickspace an \thickspace orthonormal \thickspace basis\thickspace of \thickspace R^{2n}),
\label{eq10}
\end{equation}	
$B_{n}$ and $D_{n}$ are of great importance in Particle Physics\cite{ref5}.\\ 
Let $G$ be a connected Lie group. 
A semi-simple Lie group $G$ is completely determined by the system $\Pi(G)$ of its simple roots.

\subsection{Spin Lie group and its Lie algebra}
Suppose now  $M$ is an oriented Riemannian manifold. Let $\xi$ denote a principal fibre bundle of an almost complex manifold with structural group $SO(2n)$; $n$ is a positive integer.
Let $E(\xi)$ be the total space and $B$ be the base space.  

\begin{definition}\label{def2}\cite{ref11}
	A spin structure on $\xi$ is a pair $(\psi, f)$ consisting of 
	\begin{enumerate}
		\item [(i)] a principal bundle $\psi$ over $B$ with the spin group $(Spin\thickspace (m))$ as structural group; and 
		\item [(ii)] a map $f:E(\psi) \rightarrow E(\xi)$ such that the following diagram is commutative:
	\end{enumerate}
\end{definition} 

\begin{center}
	\begin{tabular}{cccccccc}
		$E(\psi)$ &$\times$ & Spin\thickspace($m$)&$\overset{}{\longrightarrow}$& $E(\psi)$& & \\
		\multirow{3}{*}{$\big\downarrow$} & & & & &$\searrow$&\\
		& & & & & & $B$ \\
		& & & & &$\nearrow$&\\
		$E(\xi)$ &$\times$ & SO($m$)& $\longrightarrow$&$E(\xi)$& & 
	\end{tabular}
\end{center}

\begin{definition}\label{def3}\cite{ref11} 
	A spin manifold is an orientable manifold $M$ together with a spin structure on the tangent bundle of $M$.
\end{definition}

\begin{definition}\label{def4}\cite{ref4}
	A spin group is a compact dimensional Lie group.
\end{definition}


\bigskip
Lie algebras are vector spaces that are convex and thus connected.
The Lie algebra of spin particles $spin\thickspace (j)$ can be represented by classical matrices,  which make it easier to see their algebraic nature\cite{ref15}, \cite{ref17}, \cite{ref22}: 

\[ spin\thickspace(j)=\begin{cases}
higgs & j=0 ;\\
fermions & j=\mathbb{Z}\left(  \frac{1}{2}\right); \thickspace \thickspace \thickspace \thickspace \mathbb{Z} \thickspace is \thickspace odd \thickspace integer;\\ 
bosons & j=\mathbb{Z}; \thickspace \thickspace \thickspace \thickspace \thickspace \thickspace \thickspace \thickspace \mathbb{Z} \thickspace is \thickspace an \thickspace integer. \thickspace
\end{cases}\]	
The Lie algebra $sl(2n,\mathbb{C})$ can represent the fermion spin Lie algebra of elementary particles in quantum physics\cite{ref21}. As indicated in the mapping below, $\mathbb{Z}$ is an odd integer with natural numbers\cite{ref6}, \cite{ref21},\cite{ref22} $n=1,2,3, \cdots,$ :
\begin{equation*}
\xymatrix{ sl(2n,\mathbb{C}) \ar[r] & spin \thickspace \mathbb{Z}\left(  \frac{1}{2}\right) \ar[r] & fermions}.
\end{equation*}
The Lie algebra $sl(2n+1,\mathbb{C})$ can represent the boson spin Lie algebra of elementary particles. The map below gives a clear view with $\mathbb{Z}$ as an integer and natural numbers\cite{ref6}, \cite{ref21},\cite{ref22} $n=1,2,3, \cdots$.

\begin{equation*}
\xymatrix{ sl(2n+1,\mathbb{C}) \ar[r] & spin \thickspace (\mathbb{Z}) \ar[r]  & bosons}.
\end{equation*}
Parafermions and parabosons have creation  and annihilation operators  that correspond to the Dynkin's root $B_{n}$.
We can lift the results from the Lie algebra level to the Lie group level. The classical Lie algebra matrices have corresponding Lie group analogues. 
The groups $GL(n,\mathbf{C})$, $SL(n,\mathbf{C} )$, $SL(n, \mathbf{C)}$, $SU(p,q)$, $SU^{*}(2n)$,
$SU(n)$, $U(n)$, $SO(n,\mathbf{C})$, $SO(n)$, $SO^{*}(2n)$, $Sp(n, \mathbf{C})$, $Sp(n)$, $Sp(2, \mathbf{R})$, $Sp(p, q)$ are
all connected. For more details, see   \cite{ref8}.
The groups $SL(n, \mathbf{C})$ and $SU(n)$ are simply connected.
The groups $GL(n,\mathbf{R})$ and $SO(p,q)$ $(0 <p < p+q)$ have two connected components.
The group $SO(2n +1,\mathbf{C})$ is doubly connected and $SO(2n, \mathbf{C})$ is fourfold connected \cite{ref8}.
We  denote by $Spin\thickspace(J)$ the spin Lie group of a quantum spin particle as follows:

\[ \mbox{Spin} \thickspace(J)=\begin{cases}
Higgs & J=0 ;\\
Fermions & J=\mathbb{Z}\left(  \frac{1}{2}\right); \thickspace \thickspace \thickspace \thickspace \mathbb{Z} \thickspace is \thickspace odd \thickspace integer;\\ 
Bosons & J=\mathbb{Z}; \thickspace \thickspace \thickspace \thickspace \thickspace \thickspace \thickspace \mathbb{Z} \thickspace is \thickspace an \thickspace integer .\thickspace
\end{cases}\]	
The Lie group $SL(2n,\mathbb{C})$ structure can represent the fermion Spin Lie group analog\cite{ref6}, \cite{ref21},\cite{ref22}: 	
\begin{equation*}
\xymatrix{ SL(2n,\mathbb{C}) \ar[r] & Spin \thickspace \bigg(\dfrac{\mathbb{Z}}{2}\bigg) \ar[r] & Fermions,}
\end{equation*}
while the Lie group $SL(2n+1,\mathbb{C})$  represents the boson Spin Lie group analog\cite{ref6}, \cite{ref21},\cite{ref22}:
\begin{equation*}
\xymatrix{ SL(2n+1,\mathbb{C}) \ar[r] & Spin \thickspace ({\mathbb{Z}}) \ar[r] & Bosons}.
\end{equation*}	

Using the angular momentum coupling of spin particles,  one can easily observe the following\cite{ref6}, \cite{ref17}, \cite{ref21},\cite{ref22}:
For integer $n=1,2,3\cdots$ 
\begin{gather*}
\mbox{Spin} \left(\thickspace \frac{1}{2}\right)=\{ particles\thickspace spanned \thickspace by \thickspace 2 \thickspace states \thickspace with\thickspace 2\times2 \thickspace matrix\thickspace basis\},
\end{gather*}
\begin{gather*}
\mbox{Spin} \left( \thickspace \frac{2n-1}{2}\right)=\{ particles\thickspace spanned \thickspace by \thickspace 2n \thickspace states \thickspace with\thickspace (2n)\times (2n) \thickspace matrix\thickspace basis\},
\end{gather*}
\begin{gather*}
\mbox{Spin}\left( \thickspace 1\right)=\{ particles\thickspace spanned \thickspace by \thickspace 3 \thickspace states \thickspace with\thickspace 3\times3 \thickspace matrix\thickspace basis\},
\end{gather*}
\begin{gather*}
\mbox{Spin}  \left(\thickspace n\right)=\{ particles\thickspace spanned \thickspace by \thickspace 2n+1 \thickspace states \thickspace with\thickspace (2n+1)\times (2n+1) \thickspace matrix\thickspace basis\}.
\end{gather*}

Spin half odd integer particles are fermions described by Fermi-Dirac statistics and have quantum numbers described by the Pauli exclusion principle\cite{ref22}. They include the electron, proton, neutron, quarks and leptons. In particle physics, all these particles have symmetry and matrix representations. 

\subsection{SU(2) and Wigner coefficients}
We  now review some classical groups and the group theoretical approach. We  consider the same notations as in \cite{ref9}.
$SU(2)$ is the group of transformations in 2-dimensional unitary space, that is, the group of transformation leaving the form $\mid x_{1}\mid^{2}+\mid x_{2}\mid^{2}$ invariant. This group is simply compact and possesses the Lie algebra composed of the three generators $J_{1}$, $J_{2}$, $J_{3}$, obeying the commutation rule\cite{ref9}:
\begin{equation}
[J_{i},J_{j}]=ie_{ijk}J_{k}.
\label{eq11}
\end{equation}
Next, we define a set of creation and annihilation operators $a_{1}^{*}$, $a_{2}^{*}$, $a_{1}$, $a_{1},$ which obey the commutation rule
\begin{equation*}
[a_{i},a_{j}^{*}]=\delta_{ij},\thickspace i,j=1,2,
\label{eq12}
\end{equation*}
while all other commutators vanish.
The vacuum is defined by
\begin{align*}
a_{i}\mid 0 \rangle=0,
\end{align*}
and the states
\begin{align*}
\mid j,m \rangle=\dfrac{(a_{1}^{*})^{j+m}(a_{2}^{*})^{j-m}}{[(j+m)! (j-m)!]^{\frac{1}{2}}}\mid 0 \rangle,
\end{align*}
where $j$ is defined as the negative of the minimum of $m$ or positive of the maximum  of $m$.
Since $SU(2)$ is a compact group, the representation must be finite dimensional\cite{ref9}:
\begin{align*}
J_{+}=a_{1}^{*}a_{2};,\;\;\;
J_{-}=a_{2}^{*}a_{1}, \;\;\;
J_{z}=\dfrac{1}{2}(a_{1}^{*}a_{1}-a_{2}^{*}a_{2}).
\end{align*}

We can find familiar results\cite{ref9}
\begin{align*}
J_{z}|JM\rangle=M\hbar|JM\rangle,\\
J_{\pm}|JM\rangle=\hbar\sqrt{J(J+1)-M(M\pm1)}|J,M\pm1\rangle. 
\end{align*}
Furthermore, the operators $e_{i}$ obey the commutation relations \eqref{eq11}, where
\begin{align*}
e_{1}=\frac{J_{+}+J_{-}}{2},\;\;\;
e_{2}=\frac{J_{+}-J_{-}}{2i},\;\;\;
e_{3}=J_{z}.
\end{align*}

This gives a Lie algebra realization of $SU(2)$, since equation \eqref{eq11} constitutes a necessary and sufficient condition; thus we have established a mapping from the generators of the group onto the operators\cite{ref3}, \cite{ref9} $e_{i}$:
\begin{align*}
J_{i}\rightarrow e_{i} \thickspace,\thickspace i=1,2,3,
\end{align*}
which provides a representation states for the group algebra on the states $\mid j,m\rangle$.
We now derive the Wigner coefficients of the group, that is, the coefficients coupling two states
\begin{equation*}
\mid JM \rangle=\sum_{m_{1}m_{2}}C_{m_{1}m_{2}M}^{j_{1}j_{2}J}\mid j_{1},m_{1}\rangle\mid j_{2},m_{2}\rangle\delta_{m_{1}+m_{2},M}.
\end{equation*}
To do this, we consider the coupling of two spin half.
Denoting by $S_{1}$ and $S_{2}$  two spins ($\frac{1}{2}$) angular momenta, we define the total spin
\begin{equation*}
S=S_{1}+S_{2}.
\end{equation*}
Let $S_{1}^2$, $S_{1z}$, $S_{2}^2$ and $S_{2z}$ be the individual eigenstates of the eigenstates of $S^{2}$ and $S_{z}$
satisfying:
\begin{equation*}
S_{1}^2|s_{1} m_{1}\rangle=S_{1}(S_{1}+1)\hbar^2|s_{1} m_{1}\rangle;
\end{equation*}	
\begin{equation*}
S_{2}^2|s_{2} m_{2}\rangle=S_{2}(S_{2}+1)\hbar^2|s_{2} m_{2}\rangle;
\end{equation*}
\begin{equation*}
S_{1z}|s_{1} m_{1}\rangle=m\hbar|s_{1} m_{1}\rangle;
\end{equation*}
\begin{equation*}
S_{2z}|s_{2} m_{2}\rangle=m\hbar|s_{2} m_{2}\rangle.
\end{equation*}
We define the raising and lowering operators as follows \cite{ref9},\cite{ref21}:
\begin{equation*}
S_{1\pm}|s_{1}m_{1}\rangle=\hbar\sqrt{s_{1}(s_{1}+1)-m_{1}(m_{1}+1)}|s_{1},m_{1}\pm 1\rangle,
\end{equation*}
\begin{equation*}
S_{2\pm}|s_{2}m_{2}\rangle=\hbar\sqrt{s_{2}(s_{2}+1)-m_{2}(m_{2}+1)}|s_{2},m_{2}\pm 1\rangle,
\end{equation*}
and  the tensor product basis vectors as:
\begin{equation*}
|s_{1}m_{1};s_{2}m_{2}\rangle=|s_{1}m_{1}\rangle\otimes|s_{2}m_{2}\rangle,
\end{equation*}
where $m_{1}=-s_{1}...s_{1}$ and $m_{2}=-s_{2}...s_{2}.$  We seek a transformation to a set basis denoted $|SM\rangle,$ which  obeys \cite{ref3}, \cite{ref9}, \cite{ref21}:
\begin{eqnarray}
S^{2}|SM\rangle=S(S+1)\hbar^{2}|SM\rangle,
\label{eq27} \\
S_{z}|SM\rangle=M\hbar|SM\rangle,
\label{eq28}\\
S_{\pm}|SM\rangle=\hbar\sqrt{S(S+1)-M(M\pm1)}|S,M\pm1\rangle. 
\label{eq29}
\end{eqnarray}
In relation to unitary transformation
\begin{equation}
|SM\rangle=\sum_{m_{1}m_{2}}U_{m_{1}m_{2};sm}^{s_{1}s_{2}}|m_{1}m_{2}\rangle,
\label{eq30}
\end{equation}
where $U_{i,j}^{s_{1}s_{2}}$ is the $ij^{th}$ element of the unitary matrix $U^{s_{1}s_{2}}$ that transforms the basis ${|m_{1}m_{2}\rangle}$ to the basis $|SM\rangle$ \cite{ref21}.
Using the closure property of the basis ${|m_{1}m_{2}\rangle}$, 
\begin{equation}
|SM\rangle=\sum_{m_{1}m_{2}}|s_{1}s_{2}m_{1}m_{2}\rangle\langle s_{1}s_{2}m_{1}m_{2}|SM\rangle,
\label{eq31}
\end{equation}
and comparing equation \eqref{eq30} and \eqref{eq31}  lead to: 
\begin{equation*}
U_{m_{1}m_{2};sm}^{s_{1}s_{2}}\equiv\langle s_{1}s_{2}m_{1}m_{2}|sm\rangle.
\end{equation*}
The Clebsch-Gordan  coefficients  (C.G)  are obtained as: 
\begin{equation*}
U_{m_{1}m_{2};sm}^{s_{1}s_{2}}:=C_{m_{1}m_{2}m}^{s_{1}s_{2}s}.
\end{equation*} 	
The Wigner coefficients\cite{ref23} of the SU(2) group are then derived as:
\begin{align*}
C_{m_{1}m_{2}M}^{s_{1}s_{2}S}=[2S+1]^{\frac{1}{2}}(-1)^{s_{2}+m_{2}} \bigg[\dfrac{(S+s_{1}-s_{2})!(S-s_{1}+s_{2})!(s_{1}+s_{2}-S)! }{(S+s_{1}+s_{2}+1)!(s_{1}-m_{1})!(s_{1}+m_{1})!}\\ \times \dfrac{(S+M)! (S-M)!}{(s_{2}+m_{2})!(s_{2}-m_{2})!}\bigg]^{\frac{1}{2}}\sum_{k}(-1)^{k}\\ \times\dfrac{(S+s_{2}+m_{1}-k)! (s_{1}-m_{1}+k)!}{k! (S-s_{1}+s_{2}-k)! (M+S+k)! (s_{1}-s_{2}-M+K)}.
\end{align*}
Details can be found in \cite{ref3}, \cite{ref9}, \cite{ref21}.

The  paper is organized as follows.
Section II deals with the semi-simplicity of spin particle Lie structure.
In section III,  we develop the real Lie  algebra of a  spin particle. We construct the Iwasawa decomposition in section IV.  Finally, we end with some  concluding remarks in section V.
\section{Spin Semi-simplicity}	
While  investigating  the general semi-simple Lie group structure,  one can examine a similar structure in its Lie algebra. In this section, we  prove some lemmas, which  help us lift the notion of spin particle Lie algebra to the Lie group level, and prove a statement giving a clear picture of spin particles as Lie groups.  Finally, we prove a theorem on its semi-simplicity\cite{ref4}.

\begin{lemma}\label{lem1}
	Any spin particle Lie algebra admits a Clifford algebra and a spin group structure.
\end{lemma}	
	\textbf{Proof}
	Consider any \mbox{spin} $\left( j\right),$ with $j=0,\frac{1}{2},1,\cdots,$ satisfying the spin particle commutator and anticommutator relations equations ~\eqref{eq1} ~\eqref{eq2} ~\eqref{eq3} as well as the spin Lie algebra commutation bracket rule. From the equations ~\eqref{eq5} and ~\eqref{eq6} it is obvious that the Lie algebra of spin $(j)$ is a Clifford algebra. Thus,  the spin $\left( j\right)$ Lie algebra is connected and its exponential is just  
	\begin{gather*}
	\exp : spin \left( j\right)\rightarrow Spin \left(\thickspace J\right),
	\end{gather*}
	where Spin $\left(J\right)$ is the spin group. Hence,  any spin particle admits a spin group.\cqfd
\begin{lemma}\label{lem2}
	Any spin group of a spin particle admits an almost complex spin manifold and  a spin Lie group structure.
\end{lemma}
	\textbf{Proof}
	From Lemma \ref{lem1}, any spin particle admits a spin group. 
	Also, from Definition \ref{def2}, the spin group, say Spin $\left(J\right),$ has a group structure  with an almost complex manifold. Thus, from Definition \ref{def3}, the spin particle, say Spin $\left(J\right),$ with $J=0,\frac{1}{2},\cdots,$ admits a spin manifold.
	Next, we see that any spin particle has a spin group, say Spin $\left(J\right).$ Since any spin particle has a spin manifold, we observe that Spin $\left(J\right)$ is a spin group and, hence, a spin Lie group.\cqfd
\begin{proposition}\label{prop2}
	Any spin half odd integer, (resp. integer spin) Lie group,  is a fourfold cover of the compact Lie group $SO(2n)$, (resp. a double cover $SO(2n+1)$).	
\end{proposition}
	\textbf{Proof}
	The fermion quantum structure can be given as:
	\begin{equation*}
	\xymatrix{ Spin \thickspace( J )\ar[r] & Spin \thickspace (\frac{{\mathbb{Z}}}{2}}),
	\end{equation*}	
	where $\mathbb{Z}$ is odd integer. The map	
	\begin{equation*}
	\xymatrix{ SL(2n,\mathbb{C}) \ar[r] & Spin \thickspace (\frac{\mathbb{Z}}{2}) \ar[r] & SO(2n)},
	\end{equation*}
	where $SO(2n)$ is the group of all matrices, conserves the quadratic form in $\mathbb{C}^{2n}$. 
	The compact simple Lie group $SO(2n)$ is fourfold connected, and has a center, $Z(G)$, $Z_{4}$ if $n$ is odd and $Z_{2}\times Z_{2}$ if $n$ is even. Since Spin $ (\frac{\mathbb{Z}}{2})$ is a fermion with $\mathbb{Z}$ as odd integer, then,  the diagram  
	\begin{equation*}
	\begin{tabular}{ccccc}
	$SL(2n,\mathbb{C})$ & $\longrightarrow$ & $Spin\thickspace (\frac{\mathbb{Z}}{2})$ & & \\
	& & &$\searrow$ & \\
	$\uparrow$ & & $\downarrow$ & & $D_n \rightarrow fermions$\\
	& & & $\nearrow$  & \\
	$SU^{*}(2n)$ & $\longrightarrow$ & $SO(2n)$ &
	\end{tabular}
	\end{equation*}
	must commute. Thus, a fermion is a fourfold cover of $SO(2n)$.	
	Similarly,
	the boson quantum structure can be given as:
	\begin{equation*}
	\xymatrix{ Spin \thickspace (J) \ar[r] & Spin \thickspace (\mathbb{Z}}),
	\end{equation*}	
	where $\mathbb{Z}$ is an integer. The map	
	\begin{equation*}
	\xymatrix{ SL(2n+1,\mathbb{C}) \ar[r] & Spin \thickspace (\mathbb{Z}) \ar[r] & SO(2n+1)},
	\end{equation*}
	where $SO(2n+1)$ is the group of all matrices,   conserves the quadratic form in $\mathbb{C}^{2n+1}$. 
	The compact simple Lie group $SO(2n+1)$ is doubly connected and has center, $Z(G)$, $Z_{2}$. Since Spin ($\mathbb{Z}$) is a boson,  where $\mathbb{Z}$ is an integer, then the diagram   
	\begin{equation*}
	\xymatrix{ SL(2n+1,\mathbb{C}) \ar[r] \ar[rd] & Spin \thickspace (\mathbb{Z}) \ar[d] \ar[rd] \\ &  SO(2n+1) \ar[r] & B_n \rightarrow boson}
	\end{equation*}
	must commute. Thus, a boson is a double cover of $SO(2n+1).$ See \cite{ref8} for more details.\cqfd
\begin{theorem}\label{thm1}
	Any spin Lie group, Spin $\left(J\right)$ of a spin particle is:
	\begin{enumerate}
		\item [(i)] connected;
		\item [(ii)] semi-simple if and only if its simple roots are one of the Dynkin's root system $\Pi(B_{n})$ or $\Pi(D_{n})$.
	\end{enumerate}
\end{theorem}	
	\textbf{Proof}
	For (i) we
	let Spin $\left(J\right)$ be a spin Lie group with $J=0, \frac{1}{2}, 1 \cdots.$ For Spin $\left(0\right)$, Spin $\left( \frac{1}{2}\right)$, Spin $\left(1\right)$, we have, respectively,  the diagrams: 
	
	\begin{equation*}
	\xymatrix{ SL(1,\mathbb{C}) \ar[r] \ar[rd] & Spin \thickspace (0) \ar[d] \ar[rd] \\ &  SO(1) \ar[r]   & boson(Higgs),}
	\end{equation*}
	
	\begin{equation*}
	\xymatrix{ SL(2,\mathbb{C}) \ar[r] \ar[rd] & Spin \thickspace (\frac{1}{2}) \ar[d] \ar[rd] \\ &  SO(2) \ar[r]   & fermion,}
	\end{equation*}
	
	\begin{equation*}
	\xymatrix{ SL(3,\mathbb{C}) \ar[r] \ar[rd] & Spin \thickspace (1) \ar[d] \ar[rd] \\ &  SO(3) \ar[r]   & boson.}
	\end{equation*}
	From section II, the Lie groups $SO(1)$, $SO(2)$ and $SO(3)$ are connected \cite{ref8}. The results can be extended to all spin Lie groups of elementary particles as shown in Proposition \ref{prop2}. Fermions and bosons are fourfold connected and double connected, respectively.  More specifically, the spin Lie groups such as the Spin $(\frac{1}{2})$ fourfold covers the $SO(2)$ compact Lie group, while the Spin $(1)$ double covers the $SO(3)$. Next,
	for (ii), we know that
	the creation and annihilation operators generate the semi-simple Lie algebra $g$ of rank $n,$ which is a direct sum of classical Lie algebras
	\begin{gather*}
	g=B_{m_{1}}\oplus...\oplus B_{m_{k}},
	\end{gather*} 
	where $m_{1}+...+m_{k}=n.$
	Therefore, the creation and annihilation operators of spin particles generate simple Lie algebra $g$ of rank $n$ isomorphic to the classical algebra $B_{n},$ which contains a complete system $\Phi$ of roots orthogonal with respect to the Killing form \cite{ref15}. Also, from equations ~\eqref{eq1} and ~\eqref{eq2}, when we compare the bracket relation to that of the Dynkin's root $D_{n}$, $( see \thickspace equation\eqref{eq10})$, we observe that there is a correspondence.
	From Lemma \ref{lem2}, we showed that every spin group of a spin particle is a spin Lie group. We can determine the system of simple roots for the groups $\Pi(B_{n})$ and $\Pi(D_{n})$. A semi-simple Lie group $G$ is completely determined by the system $\Pi(G)$ of its simple roots \cite{ref5}. Thus, the spin Lie group of a spin particle is completely determined by the $\Pi(G)$ of its simple roots. The converse is trivial since the groups $B_{n}$ and $D_{n},$ which are $\Pi(B_{n})$ and $\Pi(D_{n})$ (Dynkin's root system), are the operators of the quantum spin particles generated by the creation and annihilation operators of rank $n$, since the spin Lie group is connected and its Lie algebra is also semi-simple. Thus, the spin Lie group is semi-simple.\cqfd
\section{Real Lie algebra of Spin particle}
The $sl(2,\mathbb{C})$ Lie algebra can be decomposed into the compact real  $su(2)$ and imaginary  $isu(2)$ forms, or $sl(2,\mathbb{R})$ and $isl(2,\mathbb{R})$. It is only natural to seek the real form of the spin half particle Lie algebra in terms of  Pauli matrices \cite{ref22}, which are $sl(2,\mathbb{C})$ matrix basis elements.

\begin{proposition}\label{prop3}
	The real Lie algebra $\mbox{spin}  \left( {\frac{1}{2}}\right)$ of  spin half particles (Spin ($\frac{1}{2})$) is given
	by  $\mbox{spin} \left({\frac{1}{2}}\right)={\{S\in M_{2}(\mathbb{R})| Tr S=0 }\}$.

\begin{enumerate}
	\item The elements ${S_{k}}=$$\frac{\hbar}{2}\left(\begin{array}{cc}
	0 & 1\\
	-1 & 0
	\end{array}\right)$, $S_{z}=\frac{\hbar}{2}\left(\begin{array}{cc}
	1 & 0\\
	0 & -1
	\end{array}\right)$, $S_{+}=\hbar \left(\begin{array}{cc}
	0 & 1\\
	0 & 0
	\end{array}\right)$ form a basis of the spin ($\frac{1}{2}$).
	\item The commutation relations   are given by :
	\item[] $[{S_{k}},S_{z}]=-\hbar S_{x}, 	[{S_{k}},S_{+}]=\hbar S_{z}, 	[{S_{z}},S_{+}]=\hbar S_{+}$.
\end{enumerate} 
\end{proposition}

	\textbf{Proof}
	Take an arbitrary angular momentum spin $(\frac{1}{2})$ with spinors $\chi=\left(\begin{array}{cc} a \\ b
	\end{array} \right)= a\chi_{\frac{1}{2}} + b \chi_{-\frac{1}{2}}$, 
	$\chi_{\frac{1}{2}}=\left(\begin{array}{cc} 1 \\ 0
	\end{array} \right)$ and 
	$\chi_{-\frac{1}{2}}=\left(\begin{array}{cc} 0 \\ 1
	\end{array} \right)$. 
	Let
	\begin{equation*}
	S_{x}=\frac{S_{+}+S_{-}}{2}
	\;\;{\mbox and} \;\;\;
	S_{y}=\frac{S_{+}-S_{-}}{2i}.
	\end{equation*}
	From the above equations \eqref{eq27} and \eqref{eq28},
	we can write $S^{2}$ and $S_{z}$ in terms of spinors. Indeed, 
	\begin{eqnarray}
	S^{2}\chi_{\frac{1}{2}}=\hbar^{2}\frac{1}{2}\bigg(\frac{1}{2}+1\bigg)\bigg|\chi_{\frac{1}{2}}\bigg>=\frac{3}{4}\hbar^{2}\chi_{\frac{1}{2}} ,\;\;\;
	S^{2}\chi_{-\frac{1}{2}}=\hbar^{2}\frac{3}{4}\chi_{-\frac{1}{2}}.
	\label{37}
	\end{eqnarray}
	From equations \eqref{37}, we can deduce
	\begin{equation*}
	S^{2}=\dfrac{3}{4}\hbar^{2}\left(\begin{array}{cc}
	1 & 0\\
	0 & 1
	\end{array}\right)=\dfrac{3}{4}\hbar^{2}I,
	\end{equation*}
	where $I=\left(\begin{array}{cc}
	1 & 0\\
	0 & 1
	\end{array}\right)$ is the identity matrix.
	Similarly,
	\begin{equation*}
	S_{z}\chi_{\frac{1}{2}}=\dfrac{\hbar}{2} \chi_{\frac{1}{2}}
	\end{equation*}
	and \\
	\begin{equation*}
	S_{z}\chi_{-\frac{1}{2}}=-\dfrac{\hbar}{2} \chi_{-\frac{1}{2}}.
	\end{equation*}
	Therefore, 
	\begin{equation*}
	S_{z}=\dfrac{\hbar}{2}\left(\begin{array}{cc}
	1 & 0\\
	0 & -1
	\end{array}\right)=\frac{\hbar}{2}\sigma_{z}.
	\end{equation*}
	By analogous computations, we get: 
	\begin{equation*}
	S_{+}=\hbar \left(\begin{array}{cc}
	0 & 1\\
	0 & 0
	\end{array}\right)=\hbar \sigma_{+}
	\end{equation*}
	and
	\begin{equation*}
	S_{-}=\hbar \left(\begin{array}{cc}
	0 & 0\\
	1 & 0
	\end{array}\right)=\hbar \sigma_{-}.
	\end{equation*}
	Similarly,
	\begin{equation}
	S_{x}=\frac{S_{+}+S_{-}}{2}=\dfrac{\hbar}{2}\left(\begin{array}{cc}
	0 & 1\\
	1 & 0
	\end{array}\right)=\frac{\hbar}{2} \sigma_{x}
	\label{eq44}
	\end{equation}
	and 
	\begin{equation}
	S_{y}=\frac{S_{+}-S_{-}}{2i}=-i S_{k}=\dfrac{-i\hbar}{2}\left(\begin{array}{cc}
	0 & 1\\
	-1 & 0
	\end{array}\right)=\frac{-i\hbar}{2}\sigma_{k}=\dfrac{\hbar}{2}\left(\begin{array}{cc}
	0 & -i\\
	i & 0
	\end{array}\right)=\frac{\hbar}{2}\sigma_{y}.
	\label{eq45}
	\end{equation}		
Defining  the bracket $[X,Y]$ of $X,Y$ by 
	\begin{eqnarray}
	[X,Y]:=XY-YX,
	\end{eqnarray}
	we observe that:
	\begin{enumerate}
		 \item [(i)] $[X,Y]\in spin \thickspace ({\frac{1}{2}}) \thickspace if \thickspace X,Y \in spin \thickspace ( {\frac{1}{2}})$;
		
		\item [(ii)] $[X,[Y,Z]]+[Y,[Z,X]]+[Z,[X,Y]]=0 \thickspace for \thickspace X,Y,Z\in spin\thickspace ({\frac{1}{2}})$;
		
		 \item [(iii)] $[X,Y]=-[Y,X] \thickspace for \thickspace X,Y\in spin\thickspace ({\frac{1}{2}}).$
	\end{enumerate}
	Thus,  the spin $({\frac{1}{2}})$ has a real Lie algebra structure on $\mathbb{R}$.
	One can easily check that
	\begin{eqnarray*}
		[{S_{k}},S_{z}]=-\hbar S_{x},\;\;\;
 	[{S_{k}},S_{+}]=\hbar S_{z}\;\;\;
{\mbox{and}}\;\;\;
[{S_{z}},S_{+}]=\hbar S_{+}.
\end{eqnarray*}			
\cqfd

\begin{lemma}\label{lem3}
	For any  $\mbox{Spin}\left(\frac{1}{2}\right)$,  there exists an orthogonal (skew symmetric) basis element $S_{k},$ which can be transformed into $SO(2),$ a compact and  rotational matrix. 
\end{lemma}
	\textbf{Proof}
	From equation ~\eqref{eq45},  we have: 
	\begin{gather*}
	S_{y}=\frac{S_{+}-S_{-}}{2i}
	=-i S_{k}
	=\dfrac{-i\hbar}{2}\left(\begin{array}{cc}
	0 & 1\\
	-1 & 0
	\end{array}\right)
=\frac{-i\hbar}{2}\sigma_{k}
=\dfrac{\hbar}{2}\left(\begin{array}{cc}
	0 & -i\\
	i & 0
	\end{array}\right)
=\frac{\hbar}{2}\sigma_{y},
	\end{gather*}
where 	$S_{k}$ is a basis of the spin half particle from Proposition \ref{prop3}.
For skew symmetric matrix,	
	\begin{gather*}
	S_{k}^{T}=-S_{k}.
	\end{gather*}
	
	Similarly, for orthogonal matrix,  we have: 
	\begin{gather*}
	S_{k}^{-1}=S_{k}^{T}.
	\end{gather*}
	Also, we see that for any $t\in \mathbb{R},$ 
	\begin{gather*}
	\exp(tS_{k})=\hbar\left(\begin{array}{cc}
	\cos \frac{t}{2} & \sin \frac{t}{2}\\
	-\sin \frac{t}{2} & \cos\frac{t}{2}
	\end{array}\right)=\hbar k_{t}.
	\end{gather*}
	When $\hbar$=1,  $\det k_{t}= 1.$ Thus, $k_{t}$ is compact.
	Next, we show that 
	$St(i)=\{g\in G| g\cdot i=i\}$ is the stabilizer of $i$. Indeed, 	
	\begin{gather*}
	g\cdot i=i\Longleftrightarrow \dfrac{ai+b}{ci+d}=i
	\Longleftrightarrow ai+b=i(ci+d)=id-c
	\Longleftrightarrow a=d \thickspace and \thickspace c=-b.
	\end{gather*}
	Then, 
	\begin{gather*}
	St(i)=\{g=\left(\begin{array}{cc}
	a & b\\
	-b & a
	\end{array}\right)\in G| a^{2}+b^{2}=1\}.
	\end{gather*}
The relation	$a^{2}+b^{2}=1$ implies  there  exists $\theta \in [0,4\pi]$ such that:  $a=\cos \frac{\theta}{2}$ and $b=\sin \frac{\theta}{2},$  where
	\begin{gather*}
	St(i)=\{g=\left(\begin{array}{cc}
	\cos \frac{\theta}{2} & \sin \frac{\theta}{2}\\
	-\sin \frac{\theta}{2} & \cos\frac{\theta}{2}
	\end{array}\right)| 0\leq \theta \leq4\pi \}=K.
	\end{gather*}
	Thus,  $K=SO(2)$ is the rotational matrix which is compact. Therefore,  $S_{k}$ is compact for $\hbar$=1.
	Hence, the proof is completed.\cqfd

\begin{remark}
	The Pauli matrices \cite{ref22}, as seen in Proposition \ref{prop3}, are just the basis of $sl(2,\mathbb{C}).$ Moreover, 
	\begin{align*}
	sl(2,\mathbb{C})= {sl(2,\mathbb{R})} \oplus i\thickspace sl(2,\mathbb{R})= {su(2)} \oplus i\thickspace su(2),
	\end{align*}
	where ${sl(2,\mathbb{R})}$ and ${su(2)}$ are the real forms of the complex group $sl(2,\mathbb{C})$ and $su(2,\mathbb{C})$ \cite{ref13}.
	Similarly,
	\begin{eqnarray*}
	\mbox{spin} \left(\frac{1}{2}\right)= {spin \left(\frac{1}{2}\right)} \oplus i \left(\mbox{spin} \thickspace\frac{1}{2}\right),
	\end{eqnarray*}
where ${spin \left(\frac{1}{2}\right)}$ is the real form of the spin half Lie algebra.
The ${spin \left(\frac{1}{2}\right)}\in sl(2,\mathbb{C})$. Thus, it is complex, and for good notation, we write ${spin \left(\frac{1}{2},\mathbb{C}\right)}\in sl(2,\mathbb{C})$. For the real form, we write ${spin \left(\frac{1}{2},\mathbb{R}\right)}\in sl(2,\mathbb{R})$.Finally,
\begin{eqnarray*}
	\mbox{spin} \left(\frac{1}{2},\mathbb{C}\right)= {spin \left(\frac{1}{2},\mathbb{R}\right)} \oplus i \left(\mbox{spin} \thickspace\frac{1}{2},\mathbb{R}\right).
\end{eqnarray*}
For simplicity, in the next section, we shall use the usual notation $spin (\frac{1}{2})$ to be the real form ${spin \left(\frac{1}{2},\mathbb{R}\right)}$ of the spin half particle.
\end{remark}

\section{Iwasawa decomposition on Lie algebra  and Lie group Levels}
Following the Iwasawa decompostion, we can uniquely decompose any semi-simple $\mbox{spin}\left(\frac{1}{2}\right)$ particle Lie algebra as follows:
\begin{equation*}
	g=\hbar k \oplus \hbar d^{\frac{1}{2}}_{t} \oplus \hbar n_{\xi},
	\label{eq47}
\end{equation*}
where $g$ is the Lie algebra of  $\mbox{Spin} \left(\frac{1}{2}\right)$  with $k_{\theta}$= \{skew symmetric $2\times 2$ matrices\}, $d^{\frac{1}{2}}_{t}$= \{$2\times 2$ real diagonal trace zero matrices\} \mbox{and} $n_{\xi}$ =\{ upper triangular $2\times 2$ matrices with zeros on the diagonal\}. 
It is just like the Iwasawa decomposition of the $sl(2,r)$ Lie algebra when the value of  $\hbar=1$  \cite{ref10}, \cite{ref13}.		

\begin{theorem}\label{thm2}
	\textbf{Iwasawa Decomposition of  Spin $(\frac{1}{2})$ particle}\cite{ref18}
	
	\begin{enumerate}
		\item [(i)] Let $\theta,\thinspace t,\thinspace\xi$ be arbitrary real numbers,
		and put $\hbar k_{\theta}=\exp(\theta {S_{k}}),$ $\hbar d^{\frac{1}{2}}_{t}=\exp( tS_{z})$, and
		$\hbar n_{\xi}=\exp(\xi S_{+})$.	
		Then, the subgroups $\hbar^{3}KDN$ of Spin $\left({\frac{1}{2}}\right)$ are defined by: $\hbar K_{\theta}=\{\hbar k_{\theta}|\theta\in R\}$,
		$\hbar D=\{\hbar d^{\frac{1}{2}}_{t}|t\in R\}$ and $\hbar N=\{\hbar n_{\xi}|\xi\in R\}$.
		We have: \\$\hbar k_{\theta}=\left(\begin{array}{cc}
		\hbar \cos\frac{\theta}{2} & \hbar \sin\frac{\theta}{2}\\
		-\hbar \sin\frac{\theta}{2} & \hbar \cos\frac{\theta}{2}
		\end{array}\right)$, $\hbar d^{\frac{1}{2}}_{t}=\left(\begin{array}{cc}
		\hbar e^{\frac{t}{2}} & 0\\
		0 & \hbar e^{-\frac{t}{2}}
		\end{array}\right)$' $\hbar n_{\xi}=\left(\begin{array}{cc}
		\hbar  & \hbar \xi\\
		0 & \hbar 
		\end{array}\right),$
		
		$\hbar K\cong\frac{\mathbb{R}}{4\pi\mathbb{Z}}\cong T,$ $\hbar D\cong\mathbb{R}$
		, $\hbar N\cong\mathbb{R}$.	
		
		\item [(ii)] Any $\mbox{spin} \left(\frac{1}{2}\right)$ particle is uniquely decomposable in the form:
		\begin{equation}
		spin \bigg(\frac{1}{2} \bigg)=\hbar^{3}\thickspace k_{\theta}d^{\frac{1}{2}}_{t}n_{\xi}=\exp\left(\theta\langle ^{s}_{m}|{S_{k}}|^{s}_{m}\rangle\right) \cdot \exp \left( t \langle _{m}^{s}|S_{z}|_{m}^{s}\rangle\right) \cdot \exp \left( \xi \langle^{s}_{m}|S_{+}|_{m}^{s}\rangle\right).
		\label{eq48}
		\end{equation}

		If $spin \thickspace ({\frac{1}{2}})=\left(\begin{array}{cc}
		a & b\\
		c & d
		\end{array}\right)\in Spin\left({\frac{1}{2}}\right)$, then,  $\theta,t,\xi$ in Theorem \ref{thm2}(i) are given by
		the relations:
		\begin{equation}
		\exp\left({i\frac{\theta}{2}}\right)=\frac{a-ic}{\hbar^{3}\sqrt{a^{2}+c^{2}}},
		\label{eq49}
		\end{equation}
		\begin{equation}
		\exp({t})=\frac{a^{2}+c^{2}}{\hbar^{6}},
		\label{eq50}
		\end{equation}
		and
		\begin{equation}
		\xi=\frac{\hbar^{6}(ab+cd)}{a^{2}+c^{2}}.
		\label{eq51}
		\end{equation}
	\end{enumerate}
\end{theorem}

\textbf{Proof}
	Since
	\begin{eqnarray}
	\hbar k_{\theta}=\exp(\theta {S_{k}})
	=\exp\left(\theta\langle ^{s}_{m}|{S_{k}}|^{s}_{m}\rangle\right)
	= \exp\left(\theta \frac{\hbar}{2}\,\left(\begin{array}{cc}
	0 & 1\\
	-1 & 0
	\end{array}\right)\right)
	=\hbar\exp\left( \frac{\theta}{2}\left(\begin{array}{cc}
	0 & 1\\
	-1 & 0
	\end{array}\right)\right)
	\end{eqnarray}
	and
	\begin{eqnarray*}
		\hbar K_{\theta}&=&\hbar \Bigg[\sum_{n=0}^{\infty}\dfrac{(-1)^{n}}{(2n)!}\bigg(\dfrac{\theta}{2}\bigg)^{2n}\cdotp I + \sum_{n=0}^{\infty}\dfrac{(-1)^{n}}{(2n+1)!}\bigg(\dfrac{\theta}{2}\bigg)^{2n+1}\cdotp {\sigma_{k}}\Bigg]\nonumber\\
		&=&\hbar \left(\begin{array}{cc}
			\cos\frac{\theta}{2} & \sin\frac{\theta}{2}\\
			-\sin\frac{\theta}{2} & \cos\frac{\theta}{2}
		\end{array}\right).
	\end{eqnarray*}
	By isomorphism $\theta\longmapsto \hbar K_{\theta},$  we obtain:  $\hbar K\cong \dfrac{R}{4\pi \mathbb{Z}}\cong T.$ Moreover,  
	\begin{eqnarray*}
		\hbar d^{\frac{1}{2}}_{t}=\exp\left(t(S_{z})\right)=\exp\left( t \langle _{m}^{s}|S_{z}|_{m}^{s}\rangle\right)
		=\sum_{n=0}^{\infty}\dfrac{1}{n!}(t\thinspace S_{z})^{n}=\hbar \sum_{n=0}^{\infty}\dfrac{1}{n!}(t\thinspace \sigma_{z})^{n}=\hbar  \left(\begin{array}{cc}
			e^{\frac{t}{2}} & 0\\
			0 & e^{-\frac{t}{2}}
		\end{array}\right).
	\end{eqnarray*}
	By isomorphism $t\longmapsto \hbar d^{\frac{1}{2}}_{t}, $ we also have: $D\cong \mathbb{R}.$
	Now,
	since $(S_{+})^{2}=0,$ 
	\begin{eqnarray*}
		\hbar n_{\xi}=\exp\left( \xi S_{+}\right)=\hbar\exp\left( \xi \thickspace \sigma_{+}\right)=\exp\left(\xi \langle^{s}_{m}|S_{+}|_{m}^{s}\rangle\right)=\hbar  \left(\begin{array}{cc}
			1 & \xi\\
			0 & 1
		\end{array}\right).
	\end{eqnarray*}
	By matrix multiplication,  we have:
	\begin{eqnarray*}
		\left(\begin{array}{cc}
			a & b\\
			c & d
		\end{array}\right)&=&\hbar^{3}\thinspace k_{\theta}d^{\frac{1}{2}}_{t}n_{\xi}\nonumber\\&=&\exp\left(\theta\langle ^{s}_{m}|{S_{k}}|^{s}_{m}\rangle\right) \cdot\exp \left(t \langle _{m}^{s}|S_{z}|_{m}^{s}\rangle\right) \cdot \exp \left( \xi \langle^{s}_{m}|S_{+}|_{m}^{s}\rangle\right)\nonumber\\
		&=&  \left(\begin{array}{cc}
			\hbar^{3}\exp\left({\frac{t}{2}}\right)cos\frac{\theta}{2} & \hbar^{3}cos\frac{\theta}{2}\exp\left({\frac{t}{2}}\right)\xi+\hbar^{3}sin\frac{\theta}{2}\exp\left({-\frac{t}{2}}\right)\\
			-\hbar^{3}\exp\left({\frac{t}{2}}\right)sin\frac{\theta}{2} & -\hbar^{3}sin\frac{\theta}{2}\exp\left({\frac{t}{2}}\right)\xi+\hbar^{3}cos\frac{\theta}{2}\exp\left({-\frac{t}{2}}\right)
		\end{array}\right).
	\end{eqnarray*}
	yielding 
	\begin{equation*}
		a=\hbar^{3} \exp\left({\frac{t}{2}}\right)cos\frac{\theta}{2},\;\;\;
		c=-\hbar^{3}\exp\left({\frac{t}{2}}\right)sin\frac{\theta}{2},
	\end{equation*}
	and		
	\[
	a-ic=\hbar^{3} \exp\left({\frac{t}{2}+i\frac{\theta}{2}}\right).
	\]
	Hence, $|a-ic|=\hbar^{3}\exp\left({\frac{t}{2}}\right),$ we  easily get equations ~\eqref{eq49} and ~\eqref{eq50}, and
	\begin{equation}
	ab+cd=\exp({t})\xi
	\end{equation}
	from which  we can clearly obtain equation ~\eqref{eq51}.\cqfd	
	

Now we know that  Spin $\left(\frac{1}{2}\right)$ is spanned by two states: $\{|\frac{1}{2} \thickspace \frac{1}{2}\rangle, |\frac{1}{2} \thickspace, -\frac{1}{2}\rangle\}$. 
From equations ~\eqref{eq27},~\eqref{eq28} and ~\eqref{eq29},  we can calculate the angular momentum for spin half integers such as $\frac{1}{2}$, $\frac{3}{2}$, $\frac{5}{2}$ and so on$\cdots$ \cite{ref21}. \\
A question arises: what can be the $n^{th}$ term of spin half integer?
From theoretical point of view, this  can be useful  in the study of particles rotational forms. We have the following statement:

\begin{theorem}\label{thm3}
	For any $\mbox{Spin}\left(\frac{2n-1}{2}\right)$ of fermions, where $n=1,2,3\cdots,$ we have:
	\begin{enumerate}
		\item [(i)]
		\begin{gather*}
			S^{2}|SM\rangle_{n}=\bigg(\frac{4n^{2}-1}{4}\bigg)\hbar^{2}|SM\rangle. 
		\end{gather*}
		\item [(ii)]
		\begin{gather*}
			S_{z}|SM\rangle_{n}=\pm\bigg(\frac{2n-k}{2}\bigg)\hbar|SM\rangle,
		\end{gather*}
		where $k\leq 2n$, and $n=1,2,3,\cdots$ with $k=1,3,5,\cdots$.
		\item [(iii)]
		The $n^{th}$ possible states of a spin half particle is given by:
		\begin{gather*}
			M_{s_{n}}=2S_{n}+1=2n,
		\end{gather*}
		where $n=1,2,3,\cdots$
		The particle is spanned by $2n$ states as below:
		\begin{gather*}
			\bigg|\bigg(\frac{2n-1}{2}\bigg) \thickspace \pm\bigg(\frac{2n-1}{2}\bigg)\bigg\rangle,\cdots, \bigg|\bigg(\frac{2n-1}{2}\bigg) \thickspace \pm(\frac{2n-k}{2}\bigg)\bigg\rangle,
		\end{gather*}
		where $k=1,3,5,\cdots,$ with $k\leq 2n.$
		\item [(iv)]
		The ladder operators act  as follows:
		\begin{gather*}
			S_{+_{n}}\bigg|\bigg(\frac{2n-1}{2}\bigg) \thickspace \bigg(\frac{2n-k}{2}\bigg)\bigg\rangle\ =\hbar\sqrt{({k-1})n-\bigg(\frac{(k-1)(k-1)}{4}\bigg)}\bigg|S,M+1\rangle,
		\end{gather*}
		
		\begin{gather*}
			S_{+_{n}}\bigg|\bigg(\frac{2n-1}{2}\bigg), \thickspace -\bigg(\frac{2n-k}{2}\bigg)\bigg\rangle=\hbar\sqrt{({k+1})n-\bigg(\frac{(k+1)(k+1)}{4}\bigg)}\bigg|S,M+1\rangle,
		\end{gather*}	
		
		\begin{gather*}
			S_{-_{n}}\bigg|\bigg(\frac{2n-1}{2}\bigg) \thickspace \bigg(\frac{2n-k}{2}\bigg)\bigg\rangle\
			=\hbar\sqrt{({k+1})n-\bigg(\frac{(k+1)(k+1)}{4}\bigg)}\bigg|S,M-1\rangle,
		\end{gather*}
		
		\begin{gather*}
			S_{-_{n}}\bigg|\bigg(\frac{2n-1}{2}\bigg) \thickspace \bigg(\frac{k-2n}{2}\bigg)\bigg\rangle\
			=\hbar\sqrt{({k-1})n-\bigg(\frac{(k-1)(k-1)}{4}\bigg)}\bigg|S,M-1\rangle.
		\end{gather*}
		
		\item [(v)]
		The ladder operators can be splitted as:
		\begin{gather*}
			S_{\pm_{n}}=S_{x_{n}}\pm S_{k_{n}}.
		\end{gather*}
	\end{enumerate}
\end{theorem}	

	\textbf{Proof}
	For the spin half integer,  we have the sequence:
	\begin{equation}
	\frac{1}{2}, \frac{3}{2}, \frac{5}{2}\cdots, \frac{2n-1}{2},
	\label{eq57}
	\end{equation}
	\\
	where $n=1,2,3,\cdots$.
	Similarly,  we can use the sequence
	\begin{equation*}
		\frac{1}{2}, \frac{3}{2}, \frac{5}{2}\cdots, \frac{2n+1}{2},
	\end{equation*}
	where $n=0,1,2,3,\cdots.$ 
	However, we will stick to that of equation ~\eqref{eq57}. 
	From equation ~\eqref{eq27}, we have:
	%
	$S^{2}|SM\rangle=S(S+1)\hbar^{2}|SM\rangle.$ 
	For the spin half, $S=\frac{2n-1}{2}$,
	we have: 
	\begin{eqnarray*}
		S^{2}|SM\rangle=\frac{2n-1}{2} \bigg(\frac{2n-1}{2}+1 \bigg)\hbar^{2}|SM\rangle .
	\end{eqnarray*}
	By simple computations,  we  arrive at:
	\begin{eqnarray*}
		S^{2}|SM\rangle_{n}=\bigg(\frac{4n^{2}-1}{4}\bigg)\hbar^{2}|SM\rangle. 
	\end{eqnarray*}
	This proves Theorem \ref{thm2}(i). 
	Next, for (ii),  we check the $n^{th}$ term for $S_{z_{n}}:$
	\begin{eqnarray*}
		S_{z_{n}}|SM\rangle=M\hbar|SM\rangle,\;\;\;
		%
		S_{z}|SM\rangle_{n}=\pm \bigg(\frac{2n-k}{2} \bigg)\hbar|SM\rangle,
	\end{eqnarray*}
	where $k\leq 2n$, and $n=1,2,3,\cdots$ with $k=1,3,5,\cdots$,
	as required.
	%
	The $n^{th}$ possible states of a spin half particle are given by:
	\begin{eqnarray*}
		M_{s_{n}}=2S_{n}+1=2 \bigg(\frac{2n-1}{2} \bigg)+1= 2n,
	\end{eqnarray*}
	where $n=1,2,3,\cdots$
	%
	For spin $({\frac{1}{2}})$,  we have ;
	\begin{eqnarray*}
		M_{s}=2n=2(1)= 2 \thickspace states, i. e.
	\end{eqnarray*}
	\begin{gather*}
		\bigg|\frac{1}{2} \thickspace \frac{1}{2}\bigg\rangle\;\;\;\mbox{and}\;\;\;\bigg|\frac{1}{2} \thickspace -\frac{1}{2}\bigg\rangle,
	\end{gather*}
	since $\frac{2n-1}{2}=\frac{2(1)-1}{2}=\frac{1}{2}.$ 
	The case $n=1$ corresponds to spin $({\frac{1}{2}})$. 	
	Thus,  for $\mbox{Spin}\left(\frac{2n-1}{2}\right)$ particles, the spin is spanned by $2n$ states.
	It is easy to check that:
	\begin{eqnarray*}
		\bigg|\bigg(\frac{2n-1}{2}\bigg) \thickspace \pm\bigg(\frac{2n-1}{2}\bigg)\bigg\rangle,\cdots, \bigg|\bigg(\frac{2n-1}{2}\bigg) \thickspace \pm\bigg(\frac{2n-k}{2}\bigg)\bigg\rangle,
	\end{eqnarray*}
	where $k=1,3,5,\cdots$ with $k\leq 2n$ as
	required by Theorem \ref{thm3}(ii).
	
	For the (iv),  we define the ladder operators for the Spin $(\frac{2n-1}{2})$:
	\begin{eqnarray*}
		S_{+_{n}}\bigg|\bigg(\frac{2n-1}{2}\bigg) \thickspace \pm\bigg(\frac{2n-k}{2}\bigg)\bigg\rangle.
	\end{eqnarray*}
	From equation ~\eqref{eq29},  we have:
	\begin{eqnarray}
	S_{\pm}|SM\rangle=\hbar\sqrt{(S\mp M)(S\pm M+1)}|S,M\pm1\rangle
	\end{eqnarray}
	and the computation provides the actions of the raising operator as given by the relations:
	\begin{align*}
		S_{+_{n}}\bigg|\bigg(\frac{2n-1}{2}\bigg) \thickspace \bigg(\frac{2n-k}{2}\bigg)\bigg\rangle\ =\hbar\sqrt{\bigg(\frac{2n-1}{2}- \frac{2n-k}{2}\bigg)\bigg(\frac{2n-1}{2}+ \frac{2n-k}{2}+1\bigg)}\bigg|S,M+1\rangle \\
		=\hbar\sqrt{\bigg(\frac{k-1}{2}\bigg)\bigg(\frac{4n-k+1}{2}\bigg)}\bigg|S,M+1\rangle=\hbar\sqrt{({k-1})n-\bigg(\frac{(k-1)(k-1)}{4}\bigg)}\bigg|S,M+1\rangle,
	\end{align*}
	and
	\begin{eqnarray*}
		S_{+_{n}}\bigg|\bigg(\frac{2n-1}{2}\bigg), \thickspace -\bigg(\frac{2n-k}{2}\bigg)\bigg\rangle&=&S_{+_{n}}\bigg|\bigg(\frac{2n-1}{2}\bigg) \thickspace \bigg(\frac{k-2n}{2}\bigg)\bigg\rangle\nonumber\\ &=&\hbar\sqrt{\bigg(\frac{2n-1}{2}- \frac{k-2n}{2}\bigg)\bigg(\frac{2n-1}{2}+ \frac{k-2n}{2}+1\bigg)}\bigg|S,M+1\rangle \nonumber\\
		&=&\hbar\sqrt{\bigg(\frac{4n-k-1}{2}\bigg)\bigg(\frac{k+1}{2}\bigg)}\bigg|S,M+1\rangle \nonumber\\ &=&\hbar\sqrt{({k+1})n-\bigg(\frac{(k+1)(k+1)}{4}\bigg)}\bigg|S,M+1\rangle.
	\end{eqnarray*}
	Similarly, it is easy to check for the lowering operator $S_{-_{n}}$ to obtain:
	\begin{align*}
		S_{-_{n}}\bigg|\bigg(\frac{2n-1}{2}\bigg) \thickspace \bigg(\frac{2n-k}{2}\bigg)\bigg\rangle\
		=\hbar\sqrt{({k+1})n-\bigg(\frac{(k+1)(k+1)}{4}\bigg)}\bigg|S,M-1\rangle
	\end{align*}
	and
	\begin{align*}
		S_{-_{n}}\bigg|\bigg(\frac{2n-1}{2}\bigg) \thickspace \bigg(\frac{k-2n}{2}\bigg)\bigg\rangle\
		=\hbar\sqrt{({k-1})n-\bigg(\frac{(k-1)(k-1)}{4}\bigg)}\bigg|S,M-1\rangle.
	\end{align*}
	For (v), we observe
	the equations ~\eqref{eq44} and ~\eqref{eq45},  and apply these operators to the spin $(\frac{2n-1}{2})$ to obtain:
	\begin{gather*}
		S_{\pm_{n}}=S_{x_{n}}\pm iS_{y_{n}}=S_{x_{n}}\pm i(-iS_{k_{n}})=S_{x_{n}}\pm S_{k_{n}}.
	\end{gather*}
	Hence, the proof is completed.\cqfd
\begin{remark}
	Note that $S_{+_{n}}^{T}=S_{-_{n}}.$ 
	From equation ~\eqref{eq44},  we have:
	\begin{eqnarray*}
		S_{x_{n}}=\frac{S_{+_{n}}+S_{-_{n}}}{2}.
	\end{eqnarray*}
	Similarly, from equation ~\eqref{eq45}, we obtain:
	\begin{eqnarray}
	S_{y_{n}}=\frac{S_{+_{n}}-S_{-_{n}}}{2i}=\frac{\hbar \sigma_{k_{n}} }{2i}=-iS_{k_{n}}.
	\label{eq76}
	\end{eqnarray}
	For $k\leq 2n$, the above ladder operators $S_{+_{n}}$ and $S_{-_{n}}$ act as:
	\begin{align*}
		S_{+_{n}}\bigg|\bigg(\frac{2n-1}{2}\bigg) \thickspace \pm\bigg(\frac{2n-k}{2}\bigg)\bigg\rangle\
	\end{align*}
	and
	\begin{align*}
		S_{-_{n}}\bigg|\bigg(\frac{2n-1}{2}\bigg) \thickspace \pm\bigg(\frac{2n-k}{2}\bigg)\bigg\rangle.
	\end{align*}
	We observe that
	\begin{align*}
		S_{+_{n}}\bigg|\bigg(\frac{2n-1}{2}\bigg) \thickspace, +\bigg(\frac{2n-k}{2}\bigg)\bigg\rangle=0
	\end{align*}
	if and only if $k=1$.	
	Similarly,
	\begin{align*}
		S_{-_{n}}\bigg|\bigg(\frac{2n-1}{2}\bigg) \thickspace \bigg(\frac{k-2n}{2}\bigg)\bigg\rangle=0,
	\end{align*}
	where $n=1,2,3,\cdots$.
\end{remark}

\begin{theorem}\label{thm4}
	For any $\mbox{Spin}\left( {\frac{2n-1}{2}}\right)$ particle spanned by $2n$ states,  there exists an orthogonal basis $S_{k_{n}}$ in the $S_{y_{n}}$ operator which can be transformed into the group $D_{n}$ of $SO(2n)$ with natural numbers $n=1,2,3,\cdots$. 
\end{theorem}
\textbf{Proof}
	From Lemma \ref{lem3},  we observe that this is true for $n=1$. For $\mbox{Spin}\left( \frac{2n-1}{2}\right)$ particle spanned by $2n$ states,
	we consider similar arguments as for Theorem \ref{thm4}, replacing the $S_{k}$ matrix by the  $n^{th}$ matrix $S_{k_{n}}$ and deducing in same manner as in  Lemma \ref{lem3} to get the above Theorem \ref{thm4}.  
	Specifically, from Theorem \ref{thm3}, there exists $S_{k_{n}}$ in the operator $S_{y_{n}}$ from equation ~\eqref{eq76}. One can check that these matrices are orthogonal and generate  SO(2n) with $n=1,2,3\cdots$. For $n=1$ we have the compact and rotational matrix SO(2) as in the  above Lemma \ref{lem3}. This achieves the proof.\cqfd
\begin{proposition}\label{prop1}\cite{ref18}
	For any element $g$ in $\mbox{Spin} \left( \frac{1}{2}\right)$ and $\theta\in\mathbb{R},$
	let
	\begin{equation*}
		gk_{\theta}=k_{g.\theta}d^{\frac{1}{2}}_{t\left(g,\theta\right)}n_{\xi\left(g,\theta\right)}
		\label{eq82}
	\end{equation*}
	be the Iwasawa decomposition of $gk_{\theta}$. If $\hbar=1$, then,  the following conditions hold, for $g,g^{\prime}$ in
	$\mbox{Spin}\left(\frac{1}{2}\right)$:
	\begin{enumerate}
		\item (i) 
		$\left(gg^{\prime}\right).\theta\equiv g.\left(g^{\prime}.\theta\right)\;\;{\mbox{mod}}\;4\pi;$ \\
		(ii) 
		$t\left(gg^{\prime},\theta\right)=t\left(g,g^{\prime}.\theta\right)+t\left(g^{\prime},\theta\right);$ \\
		(iii) 	$g.\left(\theta+2\pi\right)=g.\theta+2\pi \thickspace \;\;{\mbox{mod}}\;4\pi$ ,
		\
		$t\left(g,\theta+2\pi\right)=t\left(g,\theta\right).$
		\item 
		If $g=\left(\begin{array}{cc}
		a & b\\
		c & d
		\end{array}\right)\in \mbox{Spin} \left( \frac{1}{2}\right)$, then,  we have:
		\item [(i)] $\exp\left({i\frac{g.\theta}{2}}\right)=\dfrac{\left(a-ic\right)\cos\frac{\theta}{2}+\left(-b+id\right)\sin\frac{\theta}{2}}{|\left(a-ic\right)\cos\frac{\theta}{2}+\left(-b+id\right)\sin\frac{\theta}{2}|};$ 
		\item [(ii)] $\exp\left({t\left(g,\theta\right)}\right)=|\left(a-ic\right)\cos\frac{\theta}{2}+\left(-b+id\right)\sin\frac{\theta}{2}|^{2};$
		\item [(iii)] 
		$\xi\left(g,\theta\right)=\exp\left({-t\left(g,\theta\right)}\right)\left[\left(ab+cd\right)\cos\theta+\frac{1}{2}\left(a^{2}-b^{2}+c^{2}-d^{2}\right)\sin\theta\right];$
		\item [(iv)] 
		$d\frac{\left(g.\theta\right)}{d\theta}=\exp\left({-t\left(g,\theta\right)}\right).$	
	\end{enumerate}
\end{proposition}
\textbf{Proof}
	Now, let
	\[
	g^{\prime}k_{\theta}=k_{\theta^{\prime}}d^{\frac{1}{2}}_{t^{\prime}}n_{\xi^{\prime}},
	\]

	where $\theta^{\prime}=g^{\prime}.\theta$ , $t^{\prime}=t\left(g^{\prime},\theta\right)$ and
	$\xi^{\prime}=$$\xi\left(g^{\prime},\theta\right)$. Then,  for $g,g^{\prime}$ in
	$Spin \thickspace (\frac{1}{2}),$ we have:
	\[
	\left(gg^{\prime}\right)k_{\theta}=g(g^{\prime}k_{\theta})
	%
	=\left(gk_{\theta^{\prime}}\right)d^{\frac{1}{2}}_{t^{\prime}}n_{\xi^{\prime}}
	=k_{\theta^{\prime\prime}}d^{\frac{1}{2}}_{t^{\prime\prime}}n_{\xi^{\prime\prime}}d^{\frac{1}{2}}_{t^{\prime}}n_{\xi^{\prime}}
	=k_{\theta^{\prime\prime}}d^{\frac{1}{2}}_{t^{\prime\prime}}d^{\frac{1}{2}}_{t^{\prime}}{d^{\frac{1}{2}}_{t^{\prime}}}^{-1}n_{\xi^{\prime\prime}}d^{\frac{1}{2}}_{t^{\prime}}n_{\xi^{\prime}}
	\in KDN
	\]

	Using $\theta^{\prime\prime}=g.\theta^{\prime}$ , $t^{\prime\prime}=t\left(g,\theta^{\prime}\right)$,
	$\xi^{\prime\prime}=\xi\left(g,\theta^{\prime}\right),$ 
	because $N$ is normal in $DN,$ by the uniqueness of Iwasawa decomposition,
	we get:
	\begin{equation*}
		\left(gg^{\prime}\right).\theta\equiv\theta^{\prime\prime}\equiv g.\theta^{\prime}\equiv g.\left(g^{\prime}.\theta\right)\;\;{\mbox{mod}}\;\;4\pi
	\end{equation*}
	and
	\begin{equation*}
		t\left(gg^{\prime},\theta\right)=t^{\prime\prime}+t^{\prime}=t\left(g,g^{\prime}.\theta\right)+t\left(g^{\prime},\theta\right).
	\end{equation*}

	We note that $k_{\theta+2\pi}=-k_{\theta}$ from which  we obtain:
	\[
	gk_{\theta+2\pi}=-gk_{\theta}=-k_{g.\theta+2\pi}d^{\frac{1}{2}}_{t\left(g,\theta+2\pi\right)}n_{\xi(g,\theta+2\pi)}
	=k_{g.\theta+2\pi}d^{\frac{1}{2}}_{t\left(g,\theta+2\pi\right)}n_{\xi(g,\theta+2\pi)}.
	\]
	
	Thus,  again by uniqueness of Iwasawa decomposition, we have:
	\begin{equation*}
		g.\left(\theta+2\pi\right)=g.\theta+2\pi \thickspace \;\;{\mbox{mod}}\;\;4\pi
	\end{equation*}
	and 
	\begin{equation*}
		t\left(g,\theta+2\pi\right)=t\left(g,\theta\right).
	\end{equation*}
	We can express $g.\theta$ and $t\left(g,\theta\right)$
	as functions of coefficients of $g:$ 
	\begin{eqnarray*}
		gk_{\theta}\equiv & \left(\begin{array}{cc}
			a & b\\
			c & d
		\end{array}\right) & \left(\begin{array}{cc}
			\cos\frac{\theta}{2} & \sin\frac{\theta}{2}\\
			-\sin\frac{\theta}{2} & \cos\frac{\theta}{2}
		\end{array}\right)
		\equiv  \left(\begin{array}{cc}
			a\thinspace \cos\frac{\theta}{2}-b\thinspace \sin\frac{\theta}{2} & a\thinspace \sin\frac{\theta}{2}+b\thinspace \cos\frac{\theta}{2}\\
			c\thinspace \cos\frac{\theta}{2}-d\thinspace \sin\frac{\theta}{2} & c\thinspace \sin\frac{\theta}{2}+d\thinspace \cos\frac{\theta}{2}
		\end{array}\right).
	\end{eqnarray*}

	By direct computations as in Theorem \ref{thm2}, we get:
	\begin{equation}
	\exp\left({i\frac{g.\theta}{2}}\right)\exp\left({\frac{t\left(g,\theta\right)}{2}}\right)=\left(a-ic\right)\cos\frac{\theta}{2}+\left(-b+id\right)\sin\frac{\theta}{2},
	\label{eq86}
	\end{equation}
	\begin{eqnarray*}
	\exp\left({t\left(g,\theta\right)}\right)&=&|\left(a-ic\right)\cos\frac{\theta}{2}+\left(-b+id\right)\sin\frac{\theta}{2}|^{2}\\ \nonumber
	&=&\left(a^{2}+c^{2}\right)\cos^{2}\frac{\theta}{2}+(b^{2}+d^{2})\sin^{2}\frac{\theta}{2}-2(ab+cd)\cos\frac{\theta}{2}\sin\frac{\theta}{2}\\ \nonumber&=&\left(a^{2}+c^{2}\right)\frac{1+\cos\theta}{2}+(b^{2}+d^{2})\frac{1-\cos \theta}{2}-2(ab+cd)\sin\theta\\ \nonumber
	&=&\frac{1}{2}\left(a^{2}+b^{2}+c^{2}+d^{2}\right)+\frac{1}{2}\left(a^{2}-b^{2}+c^{2}-d^{2}\right)\cos\theta-(ab+cd)\sin\theta.
	\end{eqnarray*}
	From equation ~\eqref{eq86},  we arrive at:
	\begin{equation*}
		\exp\left({i\frac{g.\theta}{2}}\right)=\frac{\left(a-ic\right)\cos\frac{\theta}{2}+\left(-b+id\right)\sin\frac{\theta}{2}}{|\left(a-ic\right)\cos\frac{\theta}{2}+\left(-b+id\right)\sin\frac{\theta}{2}|}.
	\end{equation*}

	Differentiating both sides of equation ~\eqref{eq86} with respect
	to $\theta$, we have:
	\[
	i\frac{d\left(g.\theta\right)}{d\theta}+\frac{dt(g,\theta)}{d\theta}=\frac{\left\{ -\left(a-ic\right)\sin\frac{\theta}{2}+(-b+id)\cos\frac{\theta}{2}\right\} }{\exp\left({i\frac{g.\theta}{2}}\right)\exp\left({t\frac{\left(g,\theta\right)}{2}}\right)}.
	\]

	By equating the imaginary parts of both sides of this relation,
	we get:
	\begin{equation*}
		\frac{d\left(g.\theta\right)}{d\theta}=\frac{ad-bc}{\exp\left({t(g,\theta)}\right)}=\exp\left({-t\left(g,\theta\right)}\right).
	\end{equation*}

	Similarly,  from direct computation as in Theorem \ref{thm2}, we have: 
	\begin{equation*}
		\xi\left(g,\theta\right)=\exp\left({-t\left(g,\theta\right)}\right)\left[\left(ab+cd\right)\cos\theta+\frac{1}{2}\left(a^{2}-b^{2}+c^{2}-d^{2}\right)\sin\theta\right].
	\end{equation*}\cqfd
\begin{remark}
	The Iwasawa decomposition of the spin half particle into compact, rotational (Abelian), and nilpotent functions (subgroups) can also be performed for integer spin particles as well as for isospins.
\end{remark}

\section{Concluding remarks}
In this paper, we provided an extension of semi-simplicity of spin particle Lie algebra to the Lie group level. We showed that a spin particle Lie algebra admits a Clifford algebra, an almost complex manifold and a spin Lie group structure. We demonstrated that any spin half particle, (resp. integer spin), spin Lie group is a fourfold, (resp. double), cover of the $SO(2n),$ (resp. $SO(2n+1)$). 
We also proved that any spin Lie group of a spin particle is connected and semi-simple.  We constructed  the real Lie algebra of the $\mbox{Spin}\left(\frac{1}{2}\right)$ particle. We also performed the Iwasawa decomposition of the spin half into $KDN$. Finally, we applied the angular momentum coupling to  the $\mbox{Spin}\left(\frac{2n-1}{2}\right)$ particle and demonstrated that the orthogonal basis transforms into the $SO(2n)$ one, which is nothing but  the Dynkin's root $D_{n}$.

\begin{acknowledgements}
	This work is supported by the NLAGA-SIMONS grant and the TWAS Research Grant RGRA No. $17-1542 RG/MATHS/AF/AC\thickspace_-G-FR32400147$. The ICMPA-UNESCO chair is in partnership with Daniel lagolnitzer Foundation(DIF), France, and the Association pour la Promotion Scientifique de l'Afrique (APSA), supporting the development of Mathematical Physics in Africa.
\end{acknowledgements}



\section*{Data availability statement}
Data sharing is not applicable to this article as no new data were created or analyzed in this study.

\end{document}